# BaMF₄ (M = Mn, Co, Ni): New Electrode Materials for Hybrid Supercapacitor with Layered Polar Structure

# Ba*M*F$_4$ (*M* = Mn, Co, Ni): New Electrode Materials for Hybrid Supercapacitor with Layered Polar Structure

Shuang Zhou[1,£], Huimin Gao[1,£], Cheng Zhang[2], Jie Yang[1], Shaolong Tang[2,3], Qingyu Xu[1,3,*], Shuai Dong[1,*]

[1] School of Physics, Southeast University, Nanjing 211189, China

[2] Jiangsu Key Laboratory for Nanotechnology, Collaborative Innovation Center of Advanced Microstructures, Department of Physics, Nanjing University, Nanjing, 210093, China

[3] National Laboratory of Solid State Microstructures, Nanjing University, Nanjing 210093, China

Abstract: To pursuit high electrochemical performance of supercapacitors based on Faradaic charge-transfer with redox reaction or absorption/desorption effect, the intercalation efficiency of electrolyte ions into electrode materials is a crucial prerequisite to surpass the pure surface capacity with extra bulk contribution. Here we report layered barium transition metal fluorides, Ba*M*F$_4$ (*M* = Mn, Co, Ni) to be a series of new electrode materials applied in standard three-electrode configuration. Benefiting from the efficient immersing of electrolyte ions, these materials own prominent specific capacitance. Electrochemical characterizations demonstrate that all the Ba*M*F$_4$ electrodes show both capacitive behavior and Faradaic redox reactions in the cyclic voltammograms, and ability of charge storage by charging-discharging cycling with high cycling stability. Particularly, BaCoF$_4$ shows the the highest specific capacitance of 360 F g$^{-1}$ at current density of 0.6 A g$^{-1}$, even the particle size is far beyond nanometer scale. In addition, first principles calculations reveal the possible underlying mechanisms.

£ These authors contribute equally

*Corresponding authors: xuqingyu@seu.edu.cn(Q.X);sdong@seu.edu.cn(S.D.)

# 1. Introduction

The incremental global demand for energy together with the exhaustion of fossil fuels makes it impending to develop sustainable and renewable energy resources. Therefore, efficient and low-cost energy storage systems, such as batteries and supercapacitors, are essential to be exploited. Recent years, due to the high power densities, fast charge-discharge rate, and long cycle lifespan, supercapacitors have attracted a lot of attentions as the power resources in high power supplies, electric and hybrid vehicles, and portable electronic devices [1-4]. Nowadays, supercapacitors mostly play as the additional power alongside with batteries, while the weak specific energy prevents them to be used as standalone energy suppliers in applications.

Electrode materials are crucially important to affect the electrochemical performance, thus have been intensively explored. Typical electrode materials for supercapacitors are carbon based materials, transition metal oxides, sulfides, nitrides, and conducting polymer [5-19]. About the metal oxides, Conway's definition allows one to differentiate faradaic electrodes (eg. $Ni(OH)_2$, $Co_3O_4$) from pseudocapacitive ones (eg. $MnO_2$, $RuO_2$) [9,10,20-21]. To pursuit higher capacitance, appropriate electrode materials with large specific surface area, good conductivity, fast redox reaction, and electrochemical stability are essential to be developed to satisfy more demands of energy. For reversible redox reactions in battery-like materials, valence-variable metal cations always play a crucial role, which are frequently used in electrode materials. In addition, lower dimensional structures can provide larger specific surface area, which is beneficial for active electrode materials. In the present work, barium transition metal fluorides Ba$M$F$_4$ ($M$ = Mn, Co, Ni) with layered structure have been tested as electrode materials in standard three-electrode system.

Ba$M$F$_4$ have been dispersedly studied as multiferroics for its coexistence of pyroelectricity and antiferromagnetism [22-26]. They are isostructural: the corner shared $M$F$_6$ octahedra form the puckered sheets separated by Ba atoms, as shown in Fig. S1a (Supplementary Materials). The layered structure and valence-variable metal ions, such as Mn, Co, Ni, should benefit the diffusion of electrolyte ions into the layer intervals, and the Faradaic redox reactions near the surface of electrode materials can be highly improved, which contribute to optimize the electrochemical performance of supercapacitors. In addition, very recently, Xie *et al.* introduced ferroelectric BaTiO$_3$ into lithium-sulfur battery, which could suppress the polysulfide shuttle effect via the spontaneous polarization and thus efficiently enhance the battery performance [27]. Thus, it is natural to expect similar function in supercapacitors if ferroelectricity is introduced to electrodes. Following this expectation, the intrinsic room-temperature polarization of Ba$M$F$_4$, which doesn't exist in aforementioned electrode materials, eg. sulfides, nitrides, conducting polymers, and most of oxides, may be advantage to the charge transmission during the electrochemical progress.

## 2. Experimental and computational details

### 2.1. Fabrication of BaMF$_4$ and electrodes

Ba$M$F$_4$ ($M$ = Mn, Co, Ni) powders were synthesized using hydrothermal method, and finally resulted in pale pink, rose red, and golden yellow powders for $M$ = Mn, Co, and Ni, respectively [28-30]. To prepare the electrodes, the electroactive materials Ba$M$F$_4$ were mixed with acetylene black and polyvinylidene fluoride (PVDF) with a mass ratio of 8:1:1, and then fully milled with N-methyl-2-pyrrolidone (NMP). The cleaned slice of nickel foam substrates (1×1×0.1 cm) were coated with the mixture, dried under vacuum at 80 °C for 12 h, and then pressed into thin wafers (0.015 cm thick), further in vacuum at 80 °C for 24 h, which were used as the working electrodes.

### 2.2. Characterization

The structure of Ba$M$F$_4$ was studied by X-ray diffraction (XRD) (Rigaku Smartlab3) using a Cu K$\alpha$ radiation, and the morphology observation was performed by scanning electron microscope (SEM) (FEI Inspection F50). The electrochemical characterizations of electrodes were carried out on an electrochemical workstation (CHI660D, Chenhua Instruments, China). Both cyclic voltammetry and galvanostatic route were measured through a standard three-electrode system with 6 mol L$^{-1}$ (M) KOH solution as electrolyte. Platinum slice and saturated calomel electrode (SCE) were served as the counter electrode and reference electrode, respectively. Before measurement, the working electrode was impregnated within the electrolyte for 30 minutes to ensure that the electrode material was wet thoroughly.

*2.3. The first principles calculations*

The first-principles density function theory (DFT) calculation was performed based on the generalized gradient approximation (GGA) with Perdew-Burke-Ernzerh of potentials, as implemented in the Vienna *ab initio* Simulation Package (VASP) [31,32]. The cutoff energy of the plane-wave is 550 eV and the $K$-point mesh was $\Gamma$-centered 5×7×5. The Hubbard coefficient ($U_{eff} = U\text{-}J$) was applied on $M$'s $d$ orbitals. To fit the experimental band gaps and refer to previous DFT studies [23,33], the value of $U_{eff}$ was set as 2 eV/1.5 eV/2.5 eV for $M$ = Mn/Ni/Co, respectively. The crystal structures were relaxed until the Hellmann-Feynman forces were less than 0.01 eVÅ$^{-1}$. The experimental antiferromagnetism were adopted [25,34].

# 3. Results and discussion

Ba$M$F$_4$ ($M$ = Mn, Co, Ni) powers were synthesized via hydrothermal method. Fig. S1b shows their XRD patterns. All diffraction patterns can be well indexed to the orthorhombic structure of standard BaNiF$_4$ (PDF card NO.01-089-1836), confirming the space group of A2$_1$am [33]. From Mn to Ni, the positions of diffraction peaks shift to larger angle due to the decreasing lattice constant, in agreement with their standard data. Estimated from the positions of (020) peaks, the lattice constant along the *b*-axis (the stacking direction of layers) decrease from 15.098 Å for $BaMnF_4$ to 14.458 Å for $BaNiF_4$. No impurity phase is observed.

The morphology of Ba*M*F$_4$ particles are studied by SEM. As shown in Fig. S2, $BaMnF_4$ powders show the sheet structure, with typical lateral size in the order of 10 μm. Our previous transmission electron microscope (TEM) data confirmed that each $BaMnF_4$ sheet were grown in the ac plane, corresponding to the layered structure [28]. Although the SEM images of other two Ba*M*F$_4$ (*M* = Co and Ni) (Fig. S1b-c) show the regular shape of block in micrometer size, apparently plate-like structure can also be observed in the magnified images (insets of Fig. S1b-c). Such stratiformis morphology, due to the anisotropic crystal growth under hydrothermal condition, reflects the quasi-two-dimensional (quasi-2D) crystalline structure of Ba*M*F$_4$. This puckered (010) sheets provide spare space for fast transfer of ions, which is definitely beneficial for supercapacitors. As the research on pseudocapacitor and rechargeable battery has mostly focused on cation intercalation, such as oxide perovskite $LaMnO_3$, the anion-based charge storage mechanism presents a special novel concept of energy storage possessing high specific capacitance (SC) and affordability [35]. Noting the radius of $OH^-$ is about 1.37 Å, in Ba*M*F$_4$ the space height between adjacent layers is above 2.8 Å, which can provide innumerable and superb tunnels for ions transmission in the channel of layer space.

The electrochemical behaviors of Ba*M*F$_4$ electrodes were measured using a three-electrode configuration. The efficient loaded mass of Ba*M*F$_4$ samples is respectively 1.92 mg, 2.7 mg, and 2.25 mg for *M* = Mn, Co, and Ni corresponding to density of 0.13 g cm$^{-3}$, 0.18 g cm$^{-3}$, and 0.15 g cm$^{-3}$, respectively. Fig. 1a shows the CV's of $BaCoF_4$ electrode with the scanning rate increases from 10 mV s$^{-1}$ to 60 mV s$^{-1}$, show significant broad peaks located at around 0.1 and 0.4 V (vs. SCE), which is typically Faradaic redox reactions [36]. This CV behavior can only be seen as the signature of Faradaic electrode, rather than pseudocapacitive electrode like $MnO_2$ [11]. With the voltage range from -0.2 to 0.3V (vs. SCE), the CV curves of BaCoF$_4$ electrode was measured under various scan rates presenting rectangular CV shape typical of a electrochemical double layer capacitor (EDLC) behavior, shown in Fig. 1b, which may be on account of adsorption/desorption effect on the surface [8]. Conclusively, the Ba$M$F$_4$ electrode in our system, to some extent, can be regarded as hybrid supercapacitor. Although, this kind of hybrid electrochemical behavior always happened in composite materials such as oxide/activate carbon, Ba$M$F$_4$ family with layered structure can supply more free space and benefit for ions' absorption/desorption increasing the specific surface. Moreover, the spontaneous polarity of Ba$M$F$_4$ may also be a driving force for the ions near the particle surface,which will be discussed in detail latter. In Fig. 1a, the reduction peak moves from 0.3 to 0.4 V (vs. SCE), while the oxidization one has a shift in the range of 0.2~0.1V (vs. SCE). The potential of redox peaks are quite close to the ones of Co$_3$O$_4$ nanotubes [37]. Thus, the corresponding reactions are possibly occurred in variant valence between Co$^{2+}$ and Co$^{3+}$. For comparison, when the nickel foam was only coated by acetylene black and PVDF, there is no signal of charge accumulation (green straight-line in Fig. 1a), which demonstrates that the energy storage capability essentially originates from the fluorides.

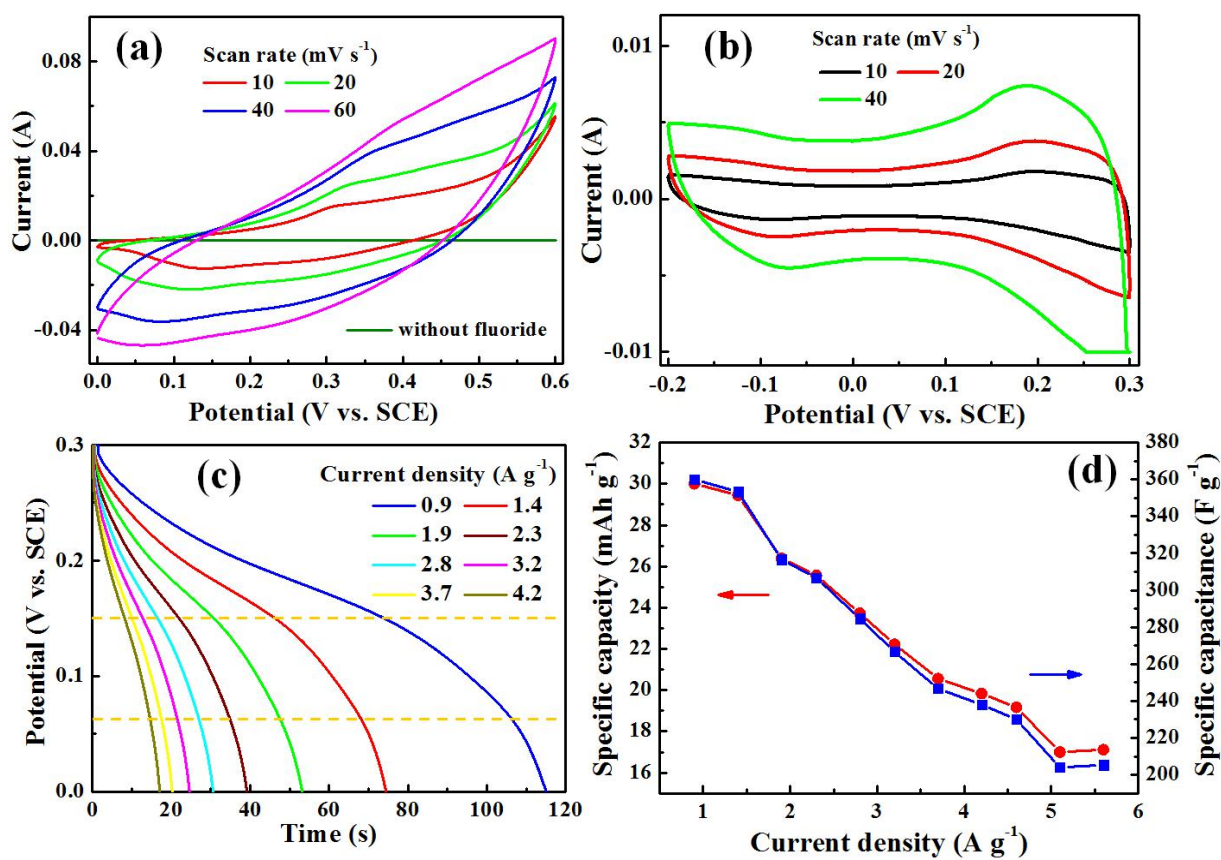

**Fig. 1.** Electrochemical characterization of BaCoF$_4$. (a) CV's measured at various scanning speed with potential window of 0-0.6 V (vs. SCE). For comparison, the horizontal green line is for the electrode without any active materials. (b) CV's measured at various scanning speed with potential window of -0.2-0.3 V (vs. SCE). (c-d) Discharge curves and calculated specific capacity (dot) and SC (square) with increasing current density.

To achieve the galvanostatic charge-discharge (GCD) curves, various current densities (from 0.9 ~ 4.2 A g$^{-1}$) were applied in the progress. Fig. 1c shows the discharge curves. Charge voltage is 0.3 V (vs. SCE) and the discharge time is gradually decreased with increasing current density. The discharge curves of BaCoF$_4$ are shown in Fig. 1c, and an inflexion point can be observed at around 1.5 V (vs. SCE), leading to unsmooth part between the two yellow dotted lines. This can be interpreted as that there are two steps of reactions (Co$^{2+}$ to Co$^{3+}$ and Co$^{4+}$) in this system [38,39]. Through the discharge curve, specific capacity based on battery materials can be calculated by $It/3.6m$, and where $I$ is constant discharge current; $t$ is discharge time; $m$ is effective mass of the active materials. On the other hand, specific capacitance of EDLC's is also evaluated by $It/m\Delta U$, where $\Delta U$ is the potential discrepancy [40]. As the current density rises from 0.6 A g$^{-1}$ to 5.5 A g$^{-1}$ in Fig. 1d, the corresponding specific capacity decreases from 30 mAh g$^{-1}$ to 16.3 mAh g$^{-1}$, corresponding to the SC of 360 F g$^{-1}$ and 220 F g$^{-1}$. The value of SC obtained here is much higher than the typical 2D structural MoS$_2$ supercapacitors (168 F g$^{-1}$ at a current density of 1 A g$^{-1}$) [41].

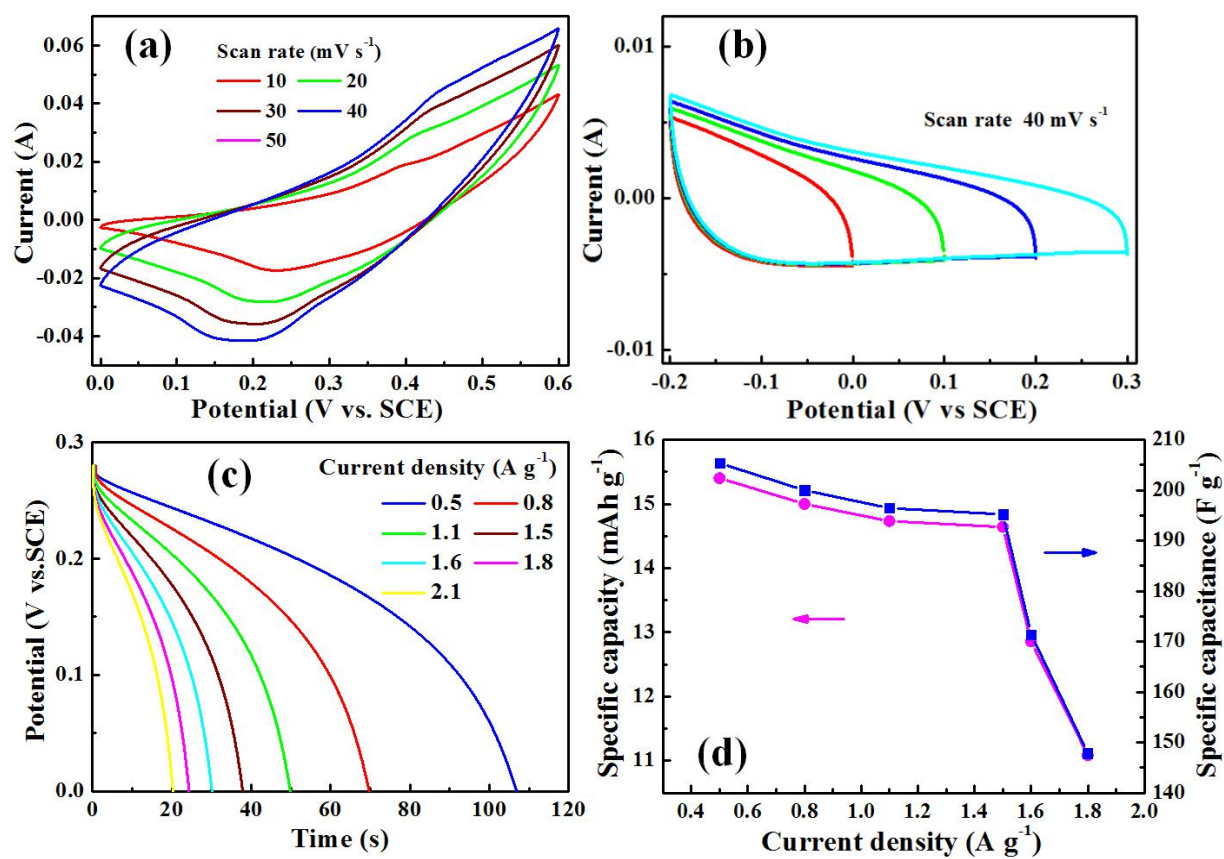

**Fig. 2.** (a-d) Electrochemical characterizations of BaMnF$_4$: (a) CV's in the potential range of 0-0.6 V (vs. SCE). (b) CV's measured by using different upper potential limits. (c-d) Discharged curves (c) and  calculated specific capacity (dot) and SC (square) with different current density (d).

The performance of BaMnF$_4$ electrode is shown in Fig. 2a. The measurement of CV's shows the proportional increase of current with various scanning rates from 10 mV s$^{-1}$ to 50 mV s$^{-1}$. Conspicuous Faradaic redox reactions are indicated by the two broad peaks located at the ranges of 0.25~0.15 V (vs. SCE) and 0.4~0.5 V (vs. SCE). Broad redox peak usually causes larger integral area of CV curve to increase SC, which is very common in MnO$_2$ [42,43]. Since Mn ions have several possible valences from +2 to +7, a few reactions are expected happening on the surface of electroactive material and leading to a relatively wide

CV like EDLC, which is helpful to improve the SC of pseudocapacitor [44]. Therefore, the reactions of Mn may occur between more than two states, which needs further verification. Fig. 2b presents the CV curves with different upper potential limits, and no redox peak was observed in the maximum potential window, revealing a capacitive behavior. GCD was carried out and the charge potential is 0.28 V  (vs. SCE). The discharge curves with different current densities are presented in Fig. 2c, and the calculated SC profile is plotted in Fig. 2d. When the current density rises from 0.5 A $g^{-1}$ to 1.5 A $g^{-1}$, the SC remains around 200 F $g^{-1}$ (15.4 mAh $g^{-1}$), which is much higher than that of carbon nanotubes/$MnO_2$ hybrid electrodes reported before [45]. For larger current density, e.g. 1.8 A $g^{-1}$, SC decreases to 150 F $g^{-1}$.

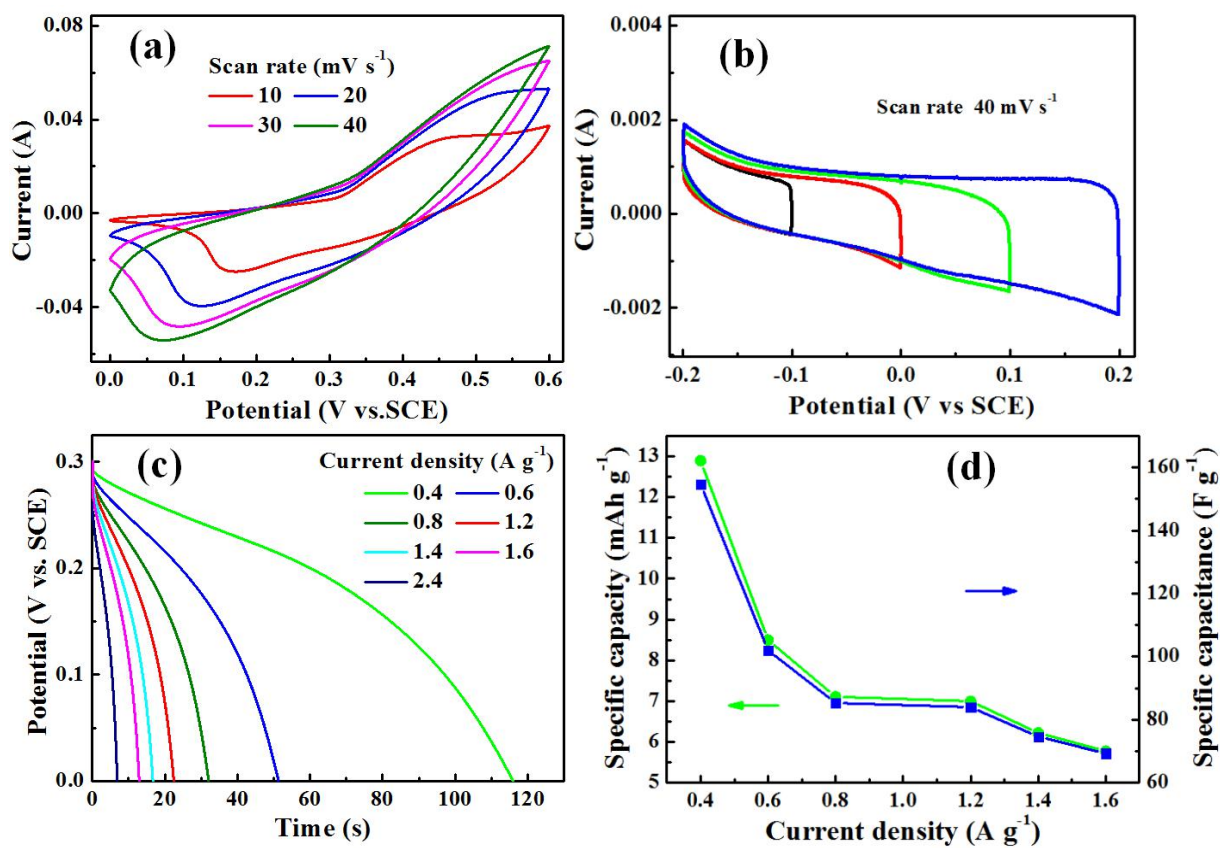

**Fig. 3** Electrochemical properties of $BaNiF_4$: (a) CV's in the potential range of 0-0.6 V (vs. SCE). (b) CV's measured by using different upper potential limits. (c) Discharged curves and (d) calculated specific capacity (dot) and SC (square) with different current density.

Besides $BaCoF_4$ and $BaMnF_4$, $BaNiF_4$ electrode also presents a promising capability of energy storage, and the characterizations are shown in Fig. 3a-d, the redox reaction peaks (Fig.

3a) are varying in the potential range of 0.15~0.5 V (vs. SCE) and 0.45~0.55 V (vs. SCE), which are similar to the values in widely studied $NiO_2$ and $Ni(OH)_2$ [46,47]. Empirically, the redox peaks are attributed to the contribution of the conversion between the common states of $Ni^{2+}$ and $Ni^{3+}$ or $Ni^{4+}$ [39,48]. As shown in Fig. 3b, CV's of $BaNiF_4$ electrode were measured with different upper potential limits, also appearing rectangular CV consistent with the results of $M$ = Mn and Co. GCD is recorded with the highest charge voltage 0.3 V (vs. SCE) and various current densities, as shown in Fig. 3c. According to the information of discharge curves, the corresponding SC is estimated in Fig. 3d. The specific capacity is 12.3 mAh $g^{-1}$, and the corresponding SC is 150 F $g^{-1}$ under the current density of 0.4 A $g^{-1}$. When the current density ascends to 1.6 A $g^{-1}$, SC decreases to 70 F $g^{-1}$.

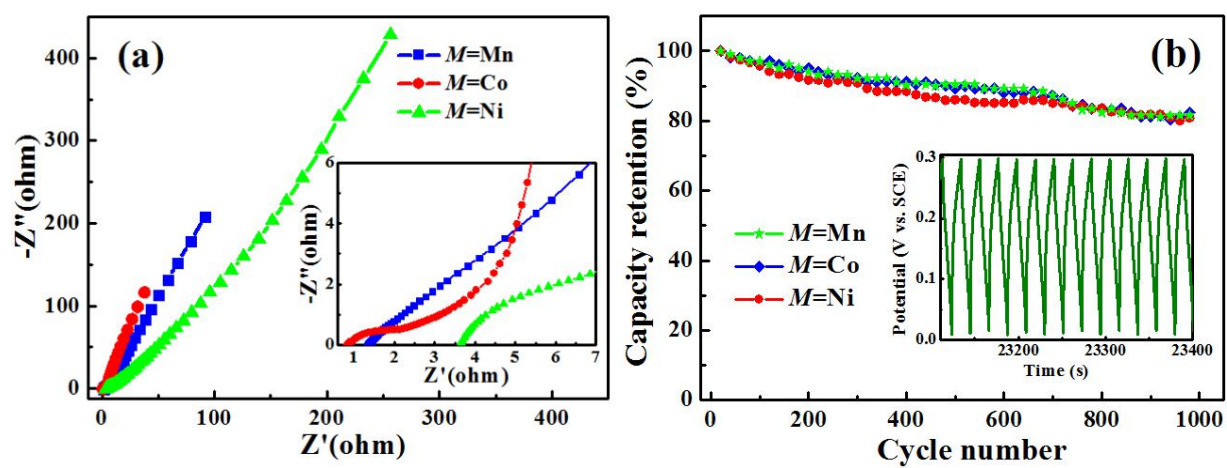

**Fig. 4.** (a) Nyquist plots of Ba$M$F$_4$ electrodes. Insert: an enlarge view. (b) Cycling performance of the three Ba$M$F$_4$ electrodes. Insert: GCD cycling of the electrode BaCoF$_4$.

In addition, the electrochemical impedance spectroscopy (EIS) was measured for the Ba$M$F$_4$ electrodes (Fig. 4a) . In the low frequency region, the slope of the curve represents the electrolyte and proton diffusion resistance [47]. The inset reveals the equivalent series resistances of the three members, i.e. BaCoF$_4$ (0.8 Ω), BaMnF$_4$ (1.4 Ω), BaNiF$_4$ (3.6 Ω), respectively. Their order of conductivity coincides with the SC's of the three members. Fig. 4b demonstrates the charge-discharge cycling stability of three electrodes, and according to the specific capacity calculation, capacity retention is n direct proportion to discharge time retention. The specific capacity of the three fluorides all maintain more than 80% after 1000 cycles, which shows a better performance than the composite materials EDLC [49-51]. In addition, the stable and symmetrical shape of charge/discharge process (inset of Fig. 3b) in the cycling test further confirms the excellent cycling performance. A. Laheaar and co-authors indicated that efficiency determined by discharge/charge durations is overestimated due to nonlinearity of the cell potential profiles for many SCs, especially those involving charge transfer reactions [40]. For non-linear GCD characteristics, the discharge energy should be calculated by integration, energy efficiency is the ratio of discharge energy to charge energy [40]. Thus the energy efficiencies of Ba$M$F$_4$ electrodes are calculated to be 61% (BaMnF$_4$), 63% (BaCoF$_4$), and 55% (BaNiF$_4$) after 1000 cycles' aging test. To have an insight into the degradation, we gained the SEM images of the loaded Ba$M$F$_4$ particles on Nickle foam. After cycling, the images show less amount of particles BaMnF (Fig. S3a-b), much fragmentation for BaNiF$_4$ (Fig. S3e-f), and almost the same feature for BaCoF$_4$ (Fig. S3c-d). The degradation maybe induced by the loose of loaded powders, fragmentation, and high concentration gradient of electrolyte.

Based on the analysis above, the probable redox reactions in the alkaline electrolyte are expressed as follows:

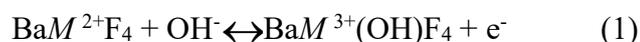
$$Ba M^{2+}F_4 + OH^- \leftrightarrow Ba M^{3+}(OH)F_4 + e^- \qquad (1)$$

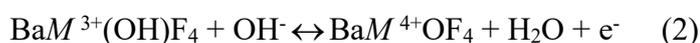
$$Ba M^{3+}(OH)F_4 + OH^- \leftrightarrow Ba M^{4+}OF_4 + H_2O + e^- \qquad (2)$$

Of course, more complicated reaction may occur beyond Eq. (1-2), especially for Mn which owns higher valences. The fluoride Ba$M$F$_4$ ($M$ = Mn, Co, Ni) has layered structure with spare space between layers, forming numerous interfaces. The space between layers plays as reservoir for anions, which can be driven in or out by external electric field or inner-build electric field (see Fig. 5a). Owing to the twist of $M$-F octahedra, Ba cations present displacement along a-axis, forming spontaneous polarization. The polarity of Ba$M$F$_4$ structure may provide an additional degree of freedom to modulate the migration of anion in more ordered process.

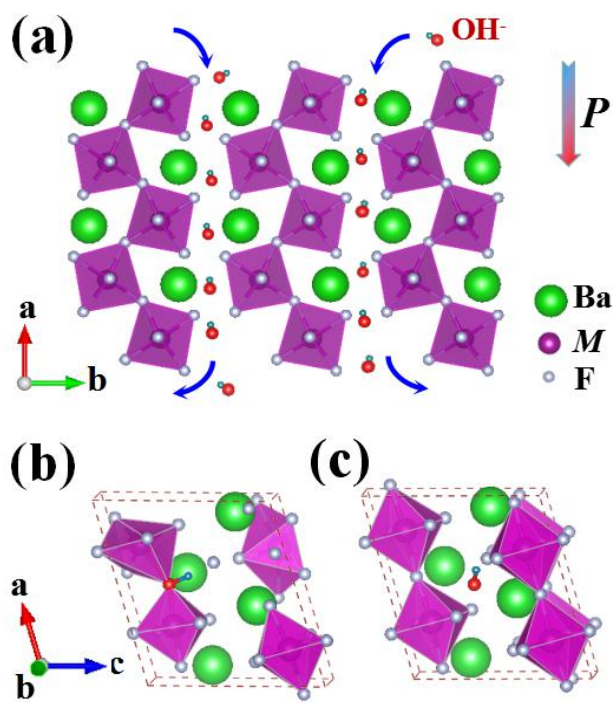

**Fig. 5.** (a) Schematic of the migration of OH group into the layer space of Ba$M$F$_4$. (b-c) Schematic of two kinds of configuration for Ba$M$F$_4$ + OH after the DFT optimization. (b) With $M$-O-$M$ bonding. (c) Non-bonding (or weakly bonding) between O and $M$ atoms.

Following this inferred model, a preliminary first-principles calculation has been performed. To simulate the reaction of Equation (2), a charge-neutral hydroxyl group (OH) is inserted into Ba$M$F$_4$ unit cell. Several positions in the layer space are tested, and then the lattice optimization is performed. The obtained final structures are shown in Fig. S4-S6. For $M$ = Mn, different initial positions of OH group lead to similar consequences that the O bonds with the neighboring Mn anion. Then the connection between adjacent octahedra is via the introduced O ion instead of original F ion. For $M$ = Ni, different initial positions of OH group also lead to similar result. But different from the Mn case, here the introduced OH group is always far from Ni, which is resident in the space between layers (Fig. S5). The Co case is between the two limits of Ni and Mn. Different initial positions of OH group generate discrepant Co-O relationship, which can be stable for bonding (Fig. S6a) or weakly-bonding (Fig. S6b-6c). The empirical evidence is that $Mn^{2+}$ and $Mn^{3+}$ are common states in oxygen octahedra structure while the stable value for Ni is divalent.

Further analysis on density of states (DOS) confirms these bonding, non-bonding, and weakly bonding (Fig. S7). In another words, $BaMnF_4$ is easy to be oxidized by OH group (the discharging process) but the opposite deoxidization process (the charging process) may be a little bit difficult. In contrast, $BaNiF_4$ is difficult to be oxidized by OH group. The circumstance of $BaCoF_4$ is just in the optimal position for the balance of oxidization/deoxidization (discharging/charging), resulting in the best performance in these three. Of course, the real situation is more complicated than our calculation, where the solution environment and applied voltage will tune the detailed balance of actions. Even though, our simplified model can give a qualitative understanding to the physical/chemical behavior of $BaMF_4$ electrodes.

Our above research demonstrated that the $BaMF_4$ family is a new series of activate material which has significant potential to be used as electrodes for supercapacitors. Comparing with other electrodes with most outstanding performance, e.g. up to 1300 F $g^{-1}$ even when the scan rate increased to 20 mV $s^{-1}$ cycles for amorphous $Ni(OH)_2$ nanospheres [7], the current performance of $BaMF_4$ is still not competitive. However, since the present work is the starting attempt to use fluorides as the electrode materials, there is enough rising space for further improvement of performance, via following routes.

First, a higher specific area of electrode material is in favor of the improvement of SC. The particle sizes of our $BaMF_4$ samples are in the order of 10 μm, which limits the charge storage. Nano-sized particles provide more contactile interfaces between electrode materials and electrolyte, which will feasibly contribute to the storage of charges, as well as the increase of conductivity [52]. Hence, to reduce the particle size of $BaMF_4$ into nanoscale will significantly improve the specific capacitance of $BaMF_4$ electrodes, according to the common sense established from previous studies on other electrode materials.Second, many previous works on transition metal oxides, e.g. $MnO_2$ and $Co_3O_4$, found great promotion of electrochemical properties after the compounding with carbon nanotubes, nanofibers, graphene, etc [53]. Combining fluorides with porous materials tend to be another method to increase the charge storage.Third, high electronic and ionic conductivity of electrode are helpful to keep the rectangular nature of CV curves, symmetry of GCD curves, and further reduce the SC losses [5]. Here the electronic conductivity of fluoride Ba$M$F$_4$ is usually not very good considering its intrinsic semiconducting properties. Thus, proper choice of better conductive binder in the fluoride electrode will benefit meliorating this weakness.The last but not the least, the selection of electrolyte is also vital for the redox reaction in pseudocapacitors. Appropriate electrolyte can conduce to a stronger ability of charge accumulation with better performance scalability.Therefore, many routes can be used to modified them to pursuit promising enhancement of electrochemical performance in Ba$M$F$_4$ pseudocapacitors.

## 4. Conclusions and perspective

A a conclusion of present work, Barium transition metal fluorides Ba$M$F$_4$ ($M$ = Mn, Co, Ni) were synthesized by hydrothermal method, and applied supercapacitor for the first time. Due to the layered structure with spare space between layers and possible benefited from its intrinsic polarity, Ba$M$F$_4$ family shows good capacity of energy storage. In standard three-electrode configuration with electrolyte of 6 M KOH, electrochemical measurements showed significant capacitive behavior and Faradaic redox reactions. For $M$ = Co, Mn, and Ni, their specific capacity are 15.4 mAh g$^{-1}$ (at 0.8 A g$^{-1}$), 30 mAh g$^{-1}$ (at 0.5 A g$^{-1}$), and 12.3 mAh g$^{-1}$ (at 0.4 A g$^{-1}$), respectively, corresponding to SC's of 360 F g$^{-1}$, 200 F g$^{-1}$, and 150 F g$^{-1}$, which are considerable values for our micrometer-sized materials. DFT analysis was performed to proof the differential performance among the three members. Among these three electrodes, $M$ = Co showed the largest SC, best conductivity, and electrochemical cycle stability. Considering the very beginning stage of Ba*M*F$_4$ as electrode, there is enough rising space to greatly improve their electrochemical performance.

## Supplementary Materials:

Structural characterization for samples and DFT results for Ba*M*F$_4$ (*M* = Mn, Co, Ni); Fig. S1 shows the Schematic structure and XRD patterns of the samples; Fig. S2 presents the morphology features; SEM images of the loading fluorides on Nickle foam (Fig. S3); Different initial OH positions and final results of structural optimization for *M* = Mn (S4), Co (S5), Ni (S6), respectively; Fig. S7 sketches DOS of BaMF$_4$ + OH, implying the bonded states for *M* - O. Computational method is described in detail.

## Acknowledgement:

This work was supported by the National Natural Science Foundation of China (51172044, 51471085, 11674055), the Natural Science Foundation of Jiangsu Province of China (BK20151400), and the open research fund of Key Laboratory of MEMS of Ministry of Education, Southeast University. Thanks for Dr. Ting Zhu's help in the revision progress.

**Supplementary Materials**

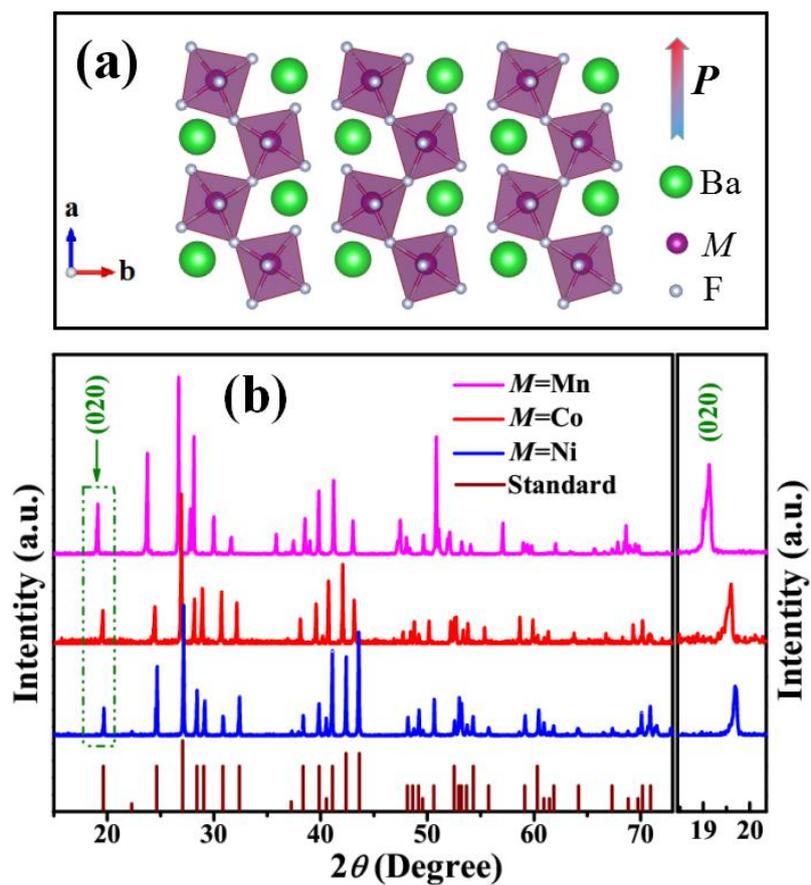

**Fig. S1.** (a) Schematic of the Ba$M$F$_4$ structure viewed along the $c$-axis. (b) X-ray diffraction patterns of Ba$M$F$_4$ showing the good crystalline and high phase purity. The standard pattern corresponds to $M$ = Ni.

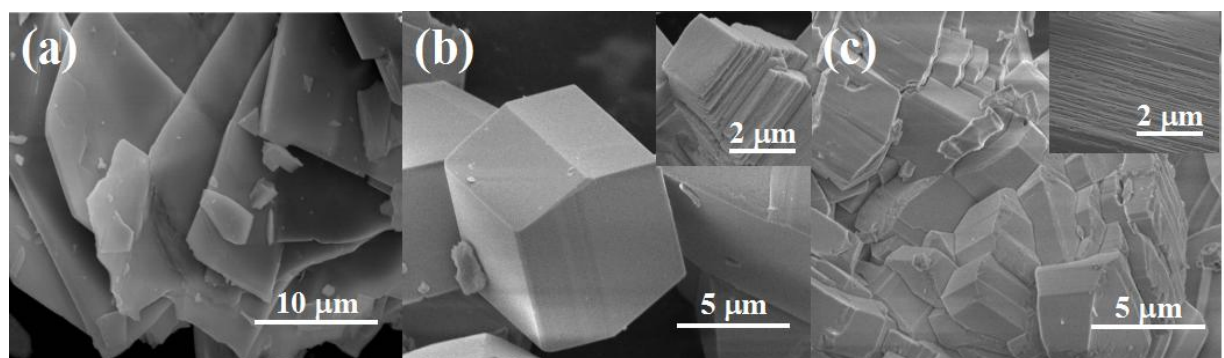

**Fig. S2.** The SEM images of Ba$M$F$_4$ ($M$ = (a) Mn, (b) Co, (c) Ni).

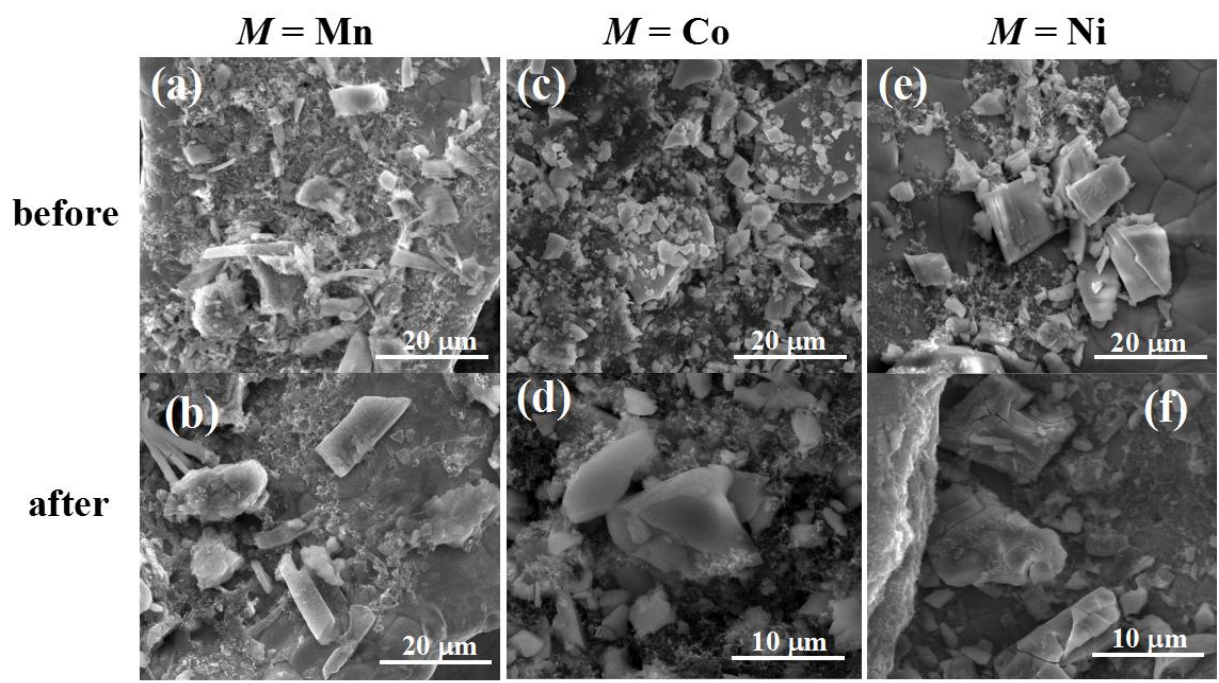

**Fig. S3** SEM images of Ba$M$F$_4$ ($M$ = (a, b) Mn, (c, d) Co, (e, f) Ni) particles attached on Nickle foam before (a, c, e) and after (b, d, f)electrochemical measurement.

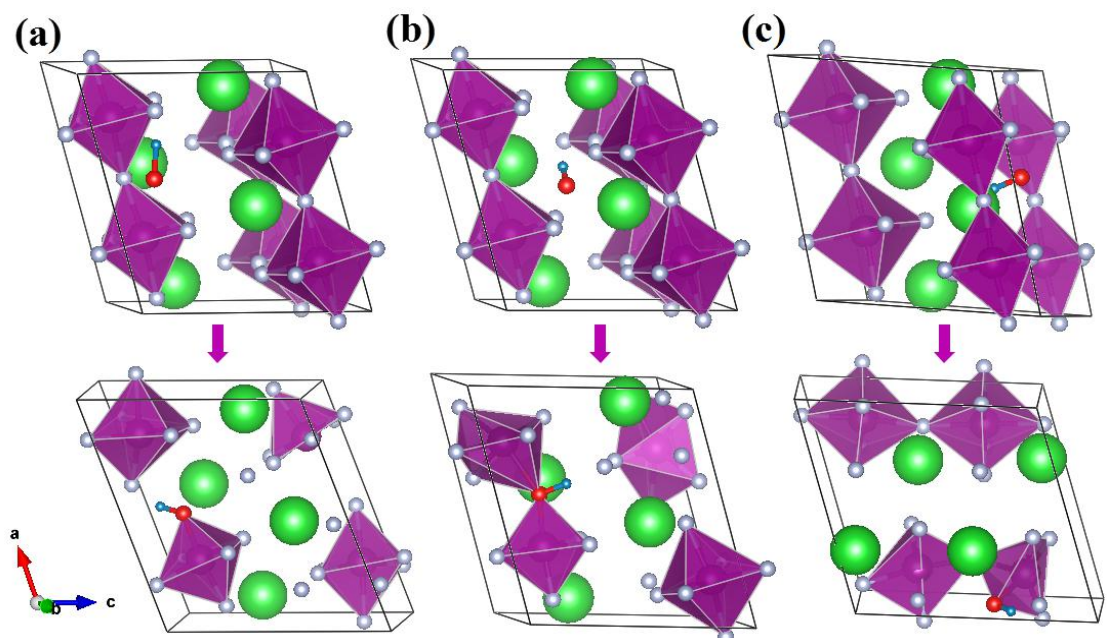

**Fig. S4.** For $M$ = Mn, different initial situations (upper panels) lead to similar final results (low panels). The O (red ball) form bonds with Mn (purple ball in polyhedron).

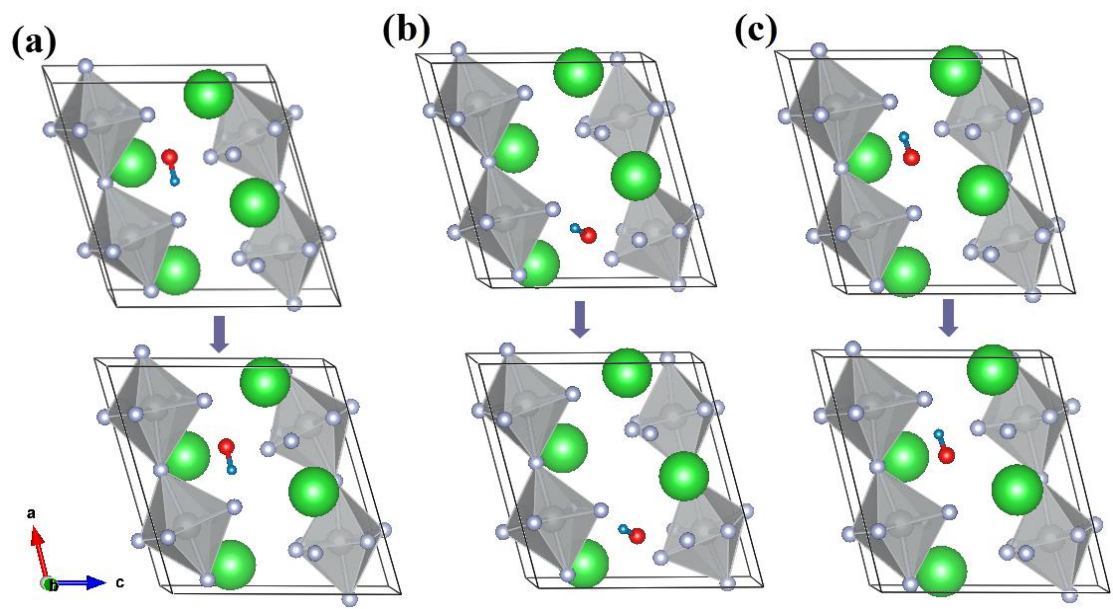

**Fig. S5.** For $M$ = Ni, the initial situations (upper panels) are hardly changed after structural optimization (lower panels), implying non-bonding between Ni and O.

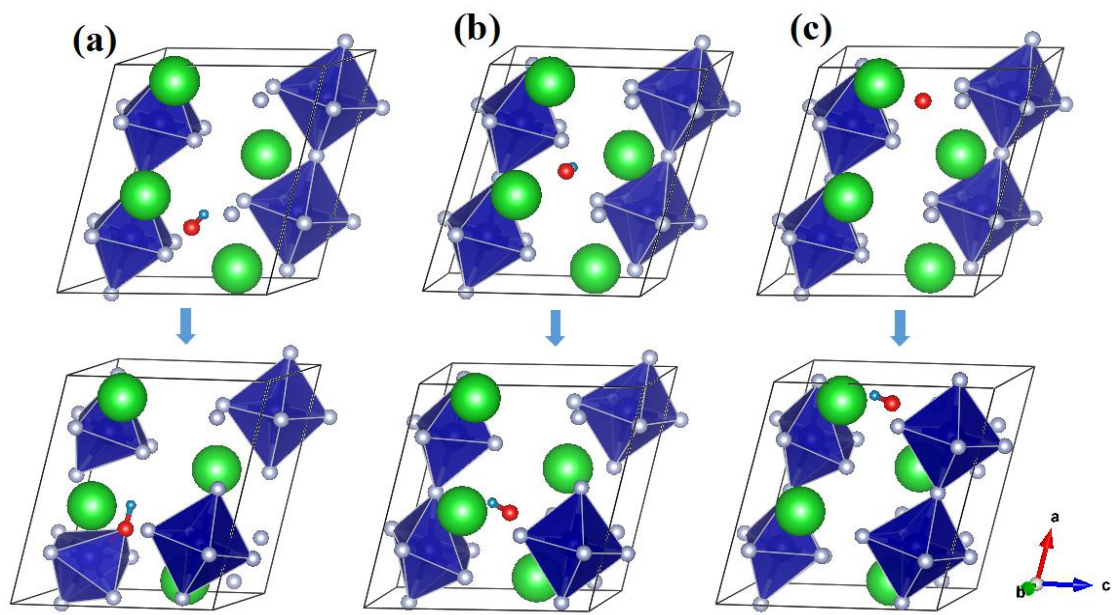

**Fig. S6.** For $M$ = Co, different initial OH positions (upper panels) cause various bonding/non-bonding conditions with Co (lower panels). (a) Co-O bonding. (b-c) weakly Co-O bonding.

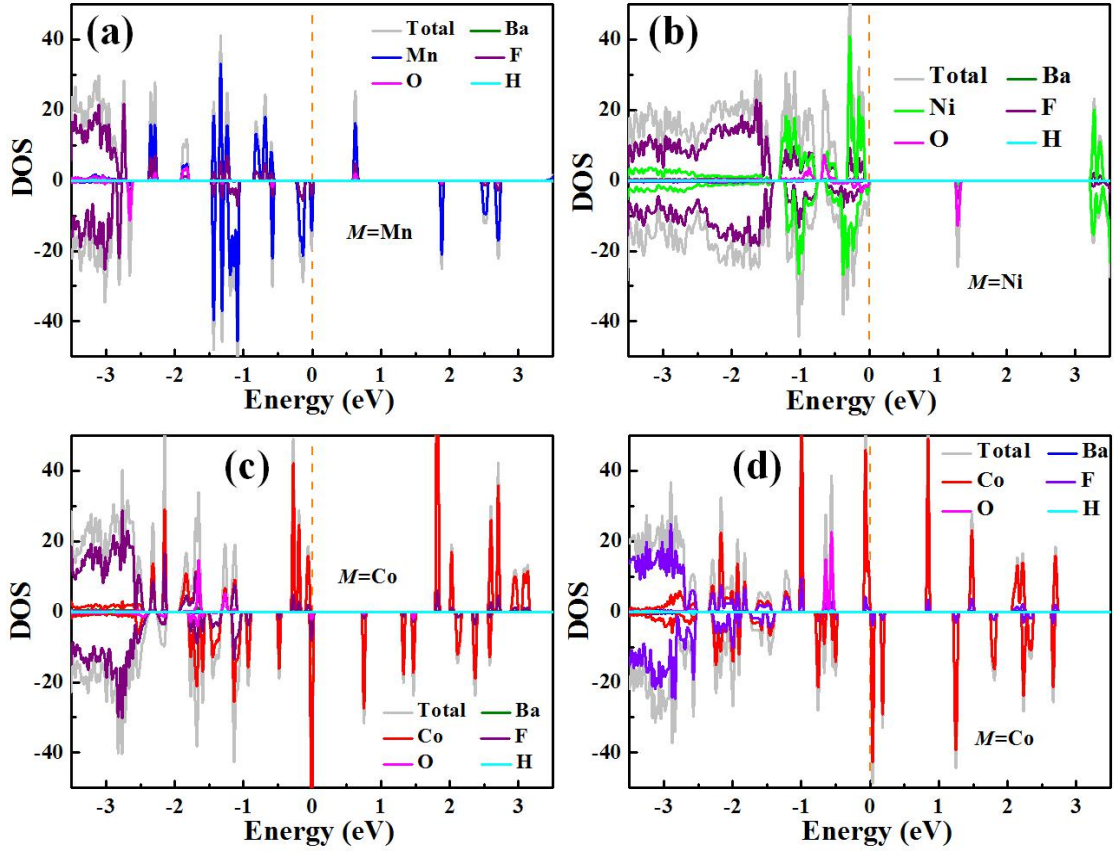

**Fig. S7.** DOS of Ba$M$F$_4$ + OH. (a) $M$ = Mn. Oxygen bands are mostly below the Fermi level, implying the bonded state. (b) $M$ = Ni. Oxygen bands are partially empty (see the state at 1.3 eV), implying the non-bonded state. For $M$ = Mn and Ni, the configurations shown in Figure S3b and S4c are selected since they own the lowest energy respectively. (c-d) $M$ = Co. (c) The Co-O bonding case (corresponding to Figure S6a with the lowest energy). Oxygen bands locate in the deep energy region (<-1 eV). (d) The weakly-bonding case (corresponding to Figure S6b with the middle energy). Oxygen bands locate in the shallow energy region (>-1 eV) but is below the Fermi level.